\documentclass{appolb}
\usepackage{epsfig}
\begin{document}
\title{Yang-Mills Thermodynamics: an Effective Theory Approach%
\thanks{Presented at XXXI Max Born Symposium and HIC for FAIR Workshop,
Wroclaw, Poland, 14-16 June 2013.}%
}
\author{Chihiro Sasaki
\address{
Frankfurt Institute for Advanced Studies,
D-60438 Frankfurt am Main,
Germany
}
}
\maketitle
\begin{abstract}
We derive the Polyakov-loop thermodynamic potential  in the 
perturbative approach to pure SU(3) Yang-Mills theory.  
The potential expressed in terms of the Polyakov loop in the fundamental
representation corresponds to that of the strong-coupling expansion,
of which the relevant coefficients of the gluon energy distribution are
specified  by characters of the SU(3) group. 
At high temperatures the derived gluon potential exhibits the 
correct asymptotic behavior, whereas at low temperatures, it 
disfavors gluons as appropriate dynamical degrees of freedom. 
In order to quantify the Yang-Mills thermodynamics
in a confined phase, we propose a hybrid approach which matches
the effective gluon potential to the one of glueballs
constrained by the QCD trace anomaly in terms of a dilaton.
We also discuss the interplay between the chromomagnetic 
and chromoelectric gluon dynamics.
\end{abstract}
\PACS{12.38.Aw, 25.75.Nq, 11.10.Wx}
  
%%%%%%%%%%%%%%%%%%%%%%%%
\section{Introduction}
%%%%%%%%%%%%%%%%%%%%%%%%%

The structure of the QCD phase diagram and thermodynamics at
finite baryon density is of crucial importance in heavy-ion
phenomenology. Due to the sign problem in lattice calculations,
a major approach to a finite density QCD is based on effective
Lagrangians possessing the same global symmetries as the underlying QCD.
The $SU(N_c)$ Yang-Mills theory has a global $Z(N_c)$ symmetry
which is dynamically broken at high temperature. This is characterized by
the Polyakov loop that plays a role of an order parameter of the $Z(N_c)$
symmetry~\cite{mclerran}. Effective models for the Polyakov loop were
suggested as a macroscopic approach to the pure gauge
theory~\cite{pisarski,pisarski:lect}. 
Their  thermodynamics is qualitatively
in agreement with that obtained in lattice gauge theories~\cite{lat:eos}.
Alternative approaches are based on the quasi-particle picture of
thermal gluons~\cite{peshier}. When gluons propagating in
a constant gluon background are considered, the quasi-particle models 
naturally merge with the Polyakov loops, that appear in the partition 
function, as characters of 
the color gauge group~\cite{turko,meisinger:2002,meisinger,kusaka,slqcd}.

In this contribution, we show that the $SU(3)$ gluon thermodynamic
potential derived from the Yang-Mills Lagrangian 
is expressed  in terms of the Polyakov loops in the fundamental 
representation. 
We  summarize  its properties and argue that at hight temperatures,  
it  exhibits the correct asymptotic behavior,
whereas at low temperatures,  it disfavors gluons~\cite{hybrid}.
We therefore suggest a hybrid approach to  Yang-Mills  thermodynamics,
which combines the  effective gluon potential with
glueballs implemented as dilaton fields.

Also, we propose an effective theory of SU(3) gluonic matter~\cite{SMR}.
The theory is constructed based on the center and scale symmetries and their
dynamical breaking, so that the interplay between color-electric and
color-magnetic gluons is included coherently. We suggest that the magnetic 
gluon condensate changes its thermal behavior qualitatively above
the critical temperature, as a consequence of matching to the 
dimensionally-reduced magnetic theories.

%%%%%%%%%%%%%%%%%%%%%%%%%%%%%%%%%%%%%%
\section{Thermodynamics of hot gluons}
%%%%%%%%%%%%%%%%%%%%%%%%%%%%%%%%%%%%%%
We start from the partition function of the pure Yang-Mills theory
\begin{equation}
Z =
\int{\mathcal D}A_\mu{\mathcal D}C{\mathcal D}\bar{C}\,
\exp\left[i\int d^4x {\mathcal L}_{\rm YM}\right]\,,
\end{equation}
with gluon $A_\mu$ and ghost $C$ fields.
Following~\cite{pisarski:lect,GPY} we employ the background field method
to evaluate  the functional integral. The gluon field is decomposed into
the background $\bar{A}_\mu$ and the quantum  $\check{A}_\mu$ fields,
\begin{equation}
A_\mu = \bar{A}_\mu + g\check{A}_\mu\,.
\end{equation}
The partition function is arranged as
\begin{eqnarray}
\ln Z
= V\int\frac{d^3 p}{(2\pi)^3}
\ln\det\left( 1 - \hat{L}_A e^{-|\vec{p}|/T}\right)
{}+ \ln M(\phi_1,\phi_2)\,,
\label{parti}
\end{eqnarray}
where $\hat{L}_A$ is the Polyakov loop matrix in the adjoint 
representation and 
the two  angular variables,  $\phi_1$ and $\phi_2$,  
represent the rank of the $SU(3)$ group.
The $M(\phi_1,\phi_2)$ is the Haar measure 
\begin{eqnarray}
\label{haar}
M(\phi_1,\phi_2)
&=&
\frac{8}{9\pi^2}\sin^2\left( \frac{\phi_1 - \phi_2}{2}\right)
\sin^2\left( \frac{2\phi_1 + \phi_2}{2}\right)
%\nonumber\\
%&&
%\times
\sin^2\left( \frac{\phi_1 + 2\phi_2}{2}\right)\,,
\end{eqnarray}
for a fixed volume $V$,  which is normalized such that
\begin{equation}
\int_0^{2\pi}\int_0^{2\pi}d\phi_1 d\phi_2 M(\phi_1,\phi_2) = 1\,.
\end{equation}
The first term in Eq.~(\ref{parti}) yields the gluon thermodynamic
potential
\begin{equation}
\Omega_g
= 2T \int\frac{d^3p}{(2\pi)^3}\mbox{tr}\ln
\left( 1 - \hat{L}_A\, e^{-E_g/T} \right)\,,
\label{omega0}
\end{equation}
where  $E_g = \sqrt{|\vec{p}|^2 + M_g^2}$ is the quasi-gluon energy 
and the effective gluon mass $M_g$ is introduced from  phenomenological
reasons.

We define the gauge invariant quantities from the Polyakov loop
matrix in the fundamental representation $\hat{L}_F$,  as
\begin{eqnarray}
\Phi = \frac{1}{3}\mbox{tr}\hat{L}_F\,,
\quad
\bar{\Phi} = \frac{1}{3}\mbox{tr}\hat{L}_F^\dagger\,.
\end{eqnarray}
Performing the trace over colors and expressing it
in terms of $\Phi$ and its conjugate $\bar\Phi$,
one arrives at
\begin{equation}
\Omega_g
= 2T \int\frac{d^3p}{(2\pi)^3}\ln
\left( 1 + \sum_{n=1}^8C_n\, e^{-nE_g/T}
\right)\,,
\label{gluon}
\end{equation}
with the coefficients $C_n$, 
\begin{eqnarray}
\label{coeff}
C_8
&=&
1\,,
\nonumber\\
C_1
&=&
C_7
= 1 - 9\bar{\Phi}\Phi\,,
\nonumber\\
C_2
&=&
C_6
= 1 - 27\bar{\Phi}\Phi
{}+ 27\left( \bar{\Phi}^3 + \Phi^3\right)\,,
\nonumber\\
C_3
&=&
C_5
= -2 + 27\bar{\Phi}\Phi
{}- 81\left( \bar{\Phi}\Phi \right)^2\,,
\nonumber\\
C_4
&=&
2\left[
-1 + 9\bar{\Phi}\Phi - 27\left( \bar{\Phi}^3 + \Phi^3\right)
{}+ 81\left( \bar{\Phi}\Phi \right)^2
\right]\,.
\end{eqnarray}
Thus, the gluon energy distribution is identified solely by the
characters of the fundamental and the conjugate representations of
the $SU(3)$ gauge group.

We introduce an effective thermodynamic potential in the large volume
limit from Eq.~(\ref{parti}) as follows:
\begin{eqnarray}
\label{full}
\Omega
&=&
\Omega_g + \Omega_\Phi + c_0\,,
\end{eqnarray}
where $\Omega_g $ is given by Eq.~(\ref{gluon}) and
the Haar measure part is found as
\begin{eqnarray}
\Omega_\Phi
&=&
-a_0T\ln\left[ 1 - 6\bar{\Phi}\Phi + 4\left( \Phi^3 + \bar{\Phi}^3\right)
{}- 3\left(\bar{\Phi}\Phi\right)^2\right]\,.
\label{dec}
\end{eqnarray}
The potential (\ref{full}) has, in general,  three free parameters;  
$a_0$,  $c_0$ and the gluon mass $M_g$.
They can be chosen e.g. to reproduce the equation of state
obtained in lattice gauge theories.
It is straightforward to see,  that the result of a non-interacting boson
gas is recovered at asymptotically high temperature.
Indeed, taking $\Phi, \bar{\Phi} \to 1$ one finds
\begin{equation}
\label{expansion}
\Omega_g(\Phi=\bar{\Phi}=1)
= 16T \int\frac{d^3p}{(2\pi)^3}
\ln\left( 1 - e^{-E_g/T} \right)\,.
\end{equation}
On the other hand, for a sufficiently large $M_g/T$, as expected near
the phase transition, one can approximate the potential as
\begin{equation}
\Omega_g
\simeq {\frac{T^2M_g^2}{\pi^2}}\sum_{n=1}^8{\frac{C_n}{n}}K_2(n\beta M_g)\,,
\label{effective}
\end{equation}
with the Bessel function $K_2(x)$.
In the quasi-particle approach, the above  result  can also be considered 
as a strong-coupling expansion,  regarding the
relation $M_g(T)=g(T)T$ with an effective gauge coupling $g(T)$.

The effective action to the  next-to-leading order of the strong coupling
expansion  is  obtained in terms of group characters as~\cite{slqcd}, 
\begin{equation}
S_{\rm eff}^{\rm (SC)}
= \lambda_{10}S_{10} + \lambda_{20}S_{20} + \lambda_{11}S_{11}
{}+ \lambda_{21}S_{21}\,,
\end{equation}
with products of  characters $S_{pq}$, specified by two integers
$p$ and $q$ counting the numbers of fundamental and conjugate
representations, and couplings $\lambda_{pq}$ being real functions of
temperature.
Making the character expansion of Eq.~(\ref{effective}),
one readily finds the correspondence between $S_{pq}$ and $C_n$ as
\begin{eqnarray}
C_{1,7} = S_{10}\,,
\quad
C_{2,6} = S_{21}\,,
\quad
C_{3,5} =S_{11}\,,
\quad
C_4 = S_{20}\,.
\end{eqnarray}

On the other hand, taking the leading contribution,  $\exp[-M_g/T]$ 
in the expansion, the ``minimal model'' is deduced with
\begin{equation}
\Omega_g
\simeq -{\mathcal F}(T,M_g)\bar{\Phi}\Phi\,,
\label{approx}
\end{equation}
where the negative sign is required for a first-order
transition~\cite{slqcd}. The function ${\mathcal F}$ can be extracted from
Eq.~(\ref{full}) and the resulting  potential is of the form 
widely used in the PNJL model~\cite{pnjl,pnjl:log,pnjl:sus,pnjl:poly}.
See also \cite{Lo}.

%%%%%%%%%%%%%%%%%%%%%%%%%%%
\section{A hybrid approach}
%%%%%%%%%%%%%%%%%%%%%%%%%%%

Although the potential (\ref{full}) describes quite well thermodynamics in 
deconfined phase, it totally fails in the confined phase.
In the confined phase,  $\langle\Phi\rangle=0$ is dynamically favored by
the ground state,   thus the $C_1=1$ term remains as the main 
contribution. Consequently
\begin{equation}
\Omega_g(\Phi=\bar{\Phi}=0)
\simeq 2T \int\frac{d^3p}{(2\pi)^3}
\ln\left( 1 + e^{-E_g/T} \right)\,.
\label{lowT}
\end{equation}
One clearly sees that $\Omega_g$ does not posses the correct sign in
front of $\exp[-E_g/T]$,  expected from the Bose-Einstein statistics.
This implies that  the entropy and the energy densities are {\it negative}.
On the other hand, if one  uses  the approximated  form (\ref{approx}), 
the pressure vanishes  at any temperature below $T_c$.
Obviously,  this is an  unphysical behavior  since there exist 
color-singlet states, i.e. glueballs, contributing to thermodynamics and
they must generate a non-vanishing pressure.

This aspect is in a striking contrast to the quark sector. 
The thermodynamic potential for  quarks  and anti-quarks
with $N_f$ flavors is obtained as~\cite{pnjl,megias}
\begin{eqnarray}
\Omega_{q+\bar q}
&=&
-2N_f T \int\frac{d^3p}{(2\pi)^3}
%\nonumber\\
%&\times&
\ln
\left[ 1 + N_c\left( \Phi + \bar{\Phi}e^{-E^+/T}\right)e^{-E^+/T}
{}+ e^{-3E^+/T}\right]
\nonumber\\
&&
{}+
\left(\mu \to -\mu\right) \,,
\label{omega:quark}
\end{eqnarray}
with $E^\pm = E_q \mp \mu$ being  the energy of a quark or anti-quark.
In the limit,  $\Phi,\bar{\Phi}\to 0$,
the one- and two-quark states are suppressed and only the three-quark 
(``baryonic'') states,  $\sim\exp(-3E^{(\pm)}/T)$, survives. This, 
on a qualitative level, is similar to confinement properties in QCD 
thermodynamics~\cite{pnjl:sus}.
One should, however,  keep in mind, that
such quark models  yield only  colored quarks being  {\it statistically}
suppressed at low temperatures.
On the other hand, unphysical thermodynamics below $T_c$ described by
the gluon sector (\ref{full}) apparently indicates,  that  gluons
are {\it physically} forbidden.
Interestingly, this property is not spoiled by the presence of quarks.
Indeed, in this case and  at  $T < T_c$  the thermodynamic potential is 
approximated as
\begin{equation}
\Omega_g + \Omega_{q+\bar{q}}
\simeq
\frac{T^2}{\pi^2}\left[
M_g^2 K_2\left(\frac{M_g}{T}\right)
{}- \frac{2N_f}{3}M_q^2 K_2\left(\frac{3M_q}{T}\right)
\right]\,.
\label{lowTapprox}
\end{equation}
Assuming that  glueballs and nucleons are made from two weakly-interacting
massive gluons and three massive quarks respectively and  putting empirical
numbers,  $M_{\rm glueball}=1.7$ GeV and $M_{\rm nucleon}=0.94$ GeV,
one finds that   $M_g = 0.85$ GeV and $M_q = 0.31$ GeV.
Substituting these mass values in Eq.~(\ref{lowTapprox}),  one still gets 
the negative entropy density 
at any temperature and   for either $N_f=2$ or $N_f=3$,  as found 
in the pure Yang-Mills theory.

The  unphysical equation of state (EoS) in confined phase can be avoided,
when gluon degrees of freedom are replaced with glueballs.
A glueball is introduced as a dilaton field $\chi$ representing the gluon
composite $\langle A_{\mu\nu}A^{\mu\nu}\rangle$,  which is responsible
for the QCD trace anomaly~\cite{schechter}. The Lagrangian is of
the standard form,
\begin{eqnarray}
\label{dilaton}
{\mathcal L}_\chi
=
\frac{1}{2}\partial_\mu\chi\partial^\mu\chi - V_\chi\,,
\quad
V_\chi
=
\frac{B}{4}\left(\frac{\chi}{\chi_0}\right)^4
\left[ \ln\left(\frac{\chi}{\chi_0}\right)^4 - 1 \right]\,,
\end{eqnarray}
with the bag constant $B$ and a dimensionful quantity $\chi_0$,
to be fixed from the vacuum energy density and the glueball mass.
One readily finds the thermodynamic potential of the glueballs as
\begin{eqnarray}
\Omega
&=&
\Omega_\chi + V_\chi + \frac{B}{4}\,,
%\nonumber\\
\quad
\Omega_\chi
=
T\int\frac{d^3p}{(2\pi)^3}\ln\left(1-e^{-E_\chi/T}\right)\,,
\nonumber\\
E_\chi
&=&
\sqrt{|\vec{p}|^2 + M_\chi^2}\,,
\quad
M_\chi^2 = \frac{\partial^2 V_\chi}{\partial\chi^2}\,,
\label{conf}
\end{eqnarray}
where a constant $B/4$ is added so that $\Omega = 0$ at zero temperature.

We propose the following hybrid approach which accounts for gluons
and glueballs degrees of freedom by combining Eqs.~(\ref{full}) and
(\ref{conf}),
\begin{equation}
\Omega = \Theta(T_c-T)\,\Omega(\chi) + \Theta(T-T_c)\,\Omega(\Phi)\,.
\label{model}
\end{equation}
For a given $M_g$,  the model parameters, $a_0$ and $c_0$, are fixed
by requiring,  that
 $\Omega(\Phi)$ yields a first-order phase transition at $T_c=270$ MeV
and that 
 $\Omega(\chi)$ and $\Omega(\Phi)$ match at $T_c$.
The resulting EoS follows general trends seen in lattice data~\cite{hybrid}.
The model can be improved further by introducing a thermal gluon mass, 
$M_g(T) \sim g(T)T$, as carried out e.g. in \cite{meisinger}.

%%%%%%%%%%%%%%%%%%%%%%%%%%%%%%%%
\section{Magnetic confinement}
%%%%%%%%%%%%%%%%%%%%%%%%%%%%%%%%

Asymptotic properties of non-abelian gauge theories at finite temperature 
are successfully captured in the quasi-particle description, which can be 
consistently calculated in the leading-order
perturbation theory~\cite{blaizot}.
However, a naive perturbative treatment in the weak coupling $g$ is spoiled
since the magnetic screening mass is dynamically generated as a ultra-soft
scale $g^2T$~\cite{Linde,GPY}. The magnetic sector
remains non-perturbative in the high temperature phase, and consequently, the
spatial string tension is non-vanishing for all temperatures~\cite{sst,sst2},
indicating certain confining properties.

This residual interaction brings apparent deviations in equations of state
(EoS) from their Stefan-Boltzmann limit at high temperature. In particular,
the interaction measure $I(T)$ is the best observable to examine dynamical
breaking of scale invariance of the Yang-Mills (YM)
Lagrangian. In lattice simulations of pure SU(3) YM theory the $I(T)/T^2T_c^2$,
with the deconfinement critical temperature $T_c$, is nearly constant in the
range $T_c < T < 5\,T_c$. This observation strongly suggests non-trivial
dynamical effects~\cite{lat:eos,latYM,dim2,CMS,Carter,fuzzy,matrix}. 
Beyond this temperature range the lattice data follow the results
from the Hard Thermal Loop (HTL) resummed perturbation theory.
Thus, a non-perturbative part in the lattice data is extracted by
subtracting the HTL contribution~\cite{latYM}.
The resultant non-perturbative part in
$I(T)/T^2T_c^2$ is {\it monotonically} decreasing, whereas the HTL result
is monotonically increasing with $T$. A plateau that arises in intermediate 
temperatures in $I(T)/T^2T_c^2$
can be therefore understood as resulting from the summation of those two 
contributions.

In \cite{SMR},
we formulate an effective theory of SU(3) gluonic matter, which accounts for
two dynamically different contributions,
the chromomagnetic and chromoelectric gluons.
In general, the dilaton
couples also to the Polyakov loop which is the order parameter of
confinement-deconfinement
phase transition and belongs to the color-electric sector.
Thus, the dilaton captures the thermodynamic properties around the critical
point $T_c$, which are related with both, the color-electric and 
color-magnetic gluons.

Thermal behavior of the magnetic gluon condensate at high temperature is found,
using the three-dimensional YM theories~\cite{AP,Nadkarni,Landsman,Kajantie},
as~\cite{Agasian}
\begin{equation}
\langle H \rangle = c_H \left(g^2(T)T\right)^4\,,
\end{equation}
with
\begin{equation}
c_H = \frac{6}{\pi}c_\sigma^2 c_m^2\,.
\end{equation}
The constants $c_\sigma$ and $c_m$ appear in $\sigma_s$ and in the
magnetic gluon mass as
\begin{equation}
\sqrt{\sigma_s(T)} = c_\sigma g^2(T)T\,,
\quad
m_g(T) = c_m g^2(T)T\,.
\end{equation}
For $SU(3)$ YM theory $c_\sigma = 0.566$~\cite{lat:eos} and
$c_m = 0.491$~\cite{3dYM}.

The  potential that mixes  the dilaton field and the Polyakov loop should be
manifestly invariant under $Z(N_c)$ and scale transformation.
For $N_c=3$, its most general form is as the following~\cite{Sannino},
\begin{equation}
V_{\rm mix} = \chi^4\left(
G_1\bar{\Phi}\Phi + G_2\left(\bar{\Phi}^3+\Phi^3\right)
{}+ G_3\left(\bar{\Phi}\Phi\right)^2 + \cdots
\right)\,,
\label{mix}
\end{equation}
with unknown coefficients $G_i$. In the following, we take only the first 
term.

At high temperature, due to the dimensional reduction, the theory
in four dimensions should match the three-dimensional YM theory.
We postulate the following matching condition,
\begin{equation}
\frac{\langle\chi\rangle}{\chi_0}
= \left(\frac{\langle H \rangle}{H_0}\right)^{1/4}\,.
\end{equation}
This transmutation from $\langle\chi\rangle \sim$ const. to
$\langle H\rangle^{1/4} \sim g^2T$ leads to an additional contribution
to the interaction measure:
\begin{equation}
\delta I
= -B\frac{\langle H\rangle}{H_0}
{}+ \left(2b_0 + \frac{b_1}{b_0}
\frac{1}{\ln\left(T/\Lambda_\sigma\right)}\right)
\frac{\langle H\rangle}{g^4(T)H_0}\,.
\end{equation}
The interaction measure normalized by $T^2T_c^2$
is monotonically decreasing even at high temperature when
no matching to the 3-dim YM is made.
The magnetic contribution generates a $T^2$ dependence.
The sum of those two contributions forms  a plateau-like behavior 
in    $I/T^2T_c^2$
at moderate temperature, $T/T_c \sim 2$-$4$.
This property appears due to the residual chromomagnetic interaction 
encoded in the dilaton, $\chi^4 \sim H$.
The resulting behavior of $I/T^2T_c^2$ with temperature
qualitatively agrees
with the latest high-precision lattice data~\cite{latYM}.
We note that a smooth switching from the dilaton to the magnetic condensate
must happen dynamically, so that thermodynamic quantities, such as the
specific heat, do not experience any irregular behavior above $T_c$.

%%%%%%%%%%%%%%%%%%
\section{Summary}
%%%%%%%%%%%%%%%%%%

We have derived the thermodynamic potential in the $SU(3)$  Yang-Mills
theory in the presence of a uniform gluon background field. 
The potential accounts for quantum statistics and reproduces 
an  ideal gas limit at high temperature.
Within the character expansion, the  one-to-one correspondence to the 
effective action in the strong-coupling expansion is obtained.
Different effective potentials used so far appear as limiting cases of
our result.

The phenomenological consequence is that
gluons are disfavored as appropriate degrees of freedom in confined phase. 
This property is in remarkable contrast to the
description of ``confinement'' within a class of chiral models with
Polyakov loops~\cite{pnjl,pnjl:poly}, where colored quarks are activated
at any temperature.

We have also presented an effective theory 
implementing the major global symmetries, the center and scale symmetries,
and their dynamical breaking. This naturally allows a mixing between the
Polyakov loop and the dilaton field. Consequently, the magnetic confinement is
effectively embedded and results in deviations of the EoS from their
Stefan-Boltzmann limit at high temperature.
Through a matching to the 3-dimensional YM theory,
the gluon condensate increases with temperature in deconfined phase.
Contrary, in the conventional treatment of the dilaton condensate,
there is a weak thermal behavior of the composite gluon in a wide range of
temperature. This suggests, that at some temperature above $T_c$, 
the gluon condensate exhibits a distinct behavior on $T$. In the
present theory this temperature is roughly estimated as $\sim 2.4\,T_c$,
compatible with $\sim 2\,T_c$ extracted from the spatial
string tension~\cite{Agasian}. Applying this idea to the interaction measure,
the role of the magnetic gluon turns out to be alternative to the HTL 
contribution.

%%%%%%%%%%%%%%%%%%%%%%%%%%%
\section*{Acknowledgments}
%%%%%%%%%%%%%%%%%%%%%%%%%%%

C.~S. acknowledges partial support by the Hessian
LOEWE initiative through the Helmholtz International
Center for FAIR (HIC for FAIR).

\end{document}